\documentclass[sigconf]{acmart}
\usepackage{amsmath}
\usepackage{graphicx}
\usepackage{textcomp}
\usepackage{graphicx}
\usepackage{array}
\usepackage{array}
\usepackage{tcolorbox}
\usepackage{booktabs}
\usepackage{placeins}
\usepackage{listings} 
\usepackage{xcolor}   
\usepackage{hyperref}
\usepackage{algorithm}
\usepackage{algpseudocode}
\usepackage{multirow}
\usepackage{makecell}

\AtBeginDocument{%
  }

\setcopyright{acmlicensed}
\copyrightyear{2026}
\acmYear{2026}
\setcopyright{cc}
\setcctype{by}
\acmConference[SAC '26]{The 41st ACM/SIGAPP Symposium on Applied Computing}{March 23--27, 2026}{Thessaloniki, Greece}
\acmBooktitle{The 41st ACM/SIGAPP Symposium on Applied Computing (SAC '26), March 23--27, 2026, Thessaloniki, Greece}
\acmDOI{10.1145/3748522.3779940}
\acmISBN{979-8-4007-2294-3/2026/03}

\begin{document}

\title[Zorya: Automated Concolic Execution of Single-Threaded Go Binaries]{Zorya: Automated Concolic Execution of\\Single-Threaded Go Binaries}


\author{Karolina Gorna}
  \affiliation{%
  \institution{Telecom Paris and Ledger Donjon}
  \city{Paris}
  \country{France}
 }
 \email{karolina.gorna@telecom-paris.fr}

 \author{Nicolas Iooss}
  \affiliation{%
  \institution{Ledger Donjon}
  \city{Zurich}
  \country{Switzerland}
  }
  \email{nicolas.iooss@ledger.com}

 \author{Yannick Seurin}
  \affiliation{%
  \institution{Ledger Donjon}
  \city{Paris}
  \country{France}
  }
  \email{yannick.seurin@ledger.com}

 \author{Rida Khatoun}
  \affiliation{%
  \institution{Telecom Paris}
  \city{Palaiseau}
  \country{France}
  }
  \email{rida.khatoun@telecom-paris.fr}

\renewcommand{\shortauthors}{Gorna et al.}

\begin{abstract}
Go's adoption in critical infrastructure intensifies the need for systematic vulnerability detection, yet existing symbolic execution tools struggle with Go binaries due to runtime complexity and scalability challenges. In this work, we build upon Zorya, a concolic execution framework that translates Go binaries to Ghidra's P-Code intermediate representation to address these challenges. We added the detection of bugs in concretely not taken paths and a multi-layer filtering mechanism to concentrate symbolic reasoning on panic-relevant paths. Evaluation on five Go vulnerabilities demonstrates that panic-reachability gating achieves 1.8–3.9× speedups when filtering 33–70\% of branches, and that Zorya detects all panics while existing tools detect at most two. Function-mode analysis proved essential for complex programs, running roughly two orders of magnitude faster than starting from \texttt{main}. This work establishes that specialized concolic execution can achieve practical vulnerability detection in language ecosystems with runtime safety checks.
\end{abstract}




\maketitle

\section{Introduction}
\label{intro}
Modern software systems increasingly rely on languages that enforce strong safety checks at runtime, yet failures still occur. Go language \cite{google_go_2025}, now prevalent in cloud infrastructure and blockchain systems, enforces many memory-safety properties through runtime checks but still exhibits panics when safety invariants are violated and can even suffer segmentation faults in the presence of \texttt{unsafe}, cgo, or low-level runtime bugs \cite{fu_golang_2024,tu_understanding_2019}. These panics are often input-dependent, triggered only when specific inputs cause nil-pointer dereferences, slice-bound violations, or nil-map operations for instance. This deterministic relationship between inputs and failures makes Go vulnerabilities well-suited to automated constraint solving. However, existing tools lack the language-specific modeling required to exploit this property.

Concolic execution, which couples concrete and symbolic execution, offers systematic vulnerability discovery by treating inputs as symbolic variables, executing programs to gather path predicates, then solving constraints to generate inputs exploring alternative paths \cite{cadar_symbolic_2013}. However, existing concolic executors face fundamental challenges when applied to Go binaries: \textbf{C1: Go-Specific Runtime Complexity.} Go binaries embed a heavyweight runtime (garbage collection, stack management, interface dispatch, slice/map/string handling) and Go-specific runtime calls and syscalls that general-purpose symbolic executors for C/C++ do not model \cite{trail_of_bits_security_2019-1}. \textbf{C2: Scalability Without Precision Loss.} Analyzing real-world Go programs requires managing vast symbolic state spaces while maintaining detection accuracy, a tension that traditional symbolic executors resolve by sacrificing either coverage or soundness.

This paper extends Zorya \cite{gorna_exposing_2025}, a binary-level concolic execution framework that lifts binaries to Ghidra \cite{nsa_ghidra_2017} P-Code. The central addition is an instantiation of classical target-directed reachability for Go panic functions, which we term \emph{panic-gated exploration}. For each concrete starting state, Zorya restricts symbolic reasoning to branches that may reach known panic locations, using a conservative backward reachability pre-analysis over the control-flow graph instead of exploring all execution paths. This specialization of reachability to Go's panic mechanism, combined with additional filtering mechanisms, lays the foundation for scalable concolic analysis of large Go binaries and, by extension, other compiled languages.

\textbf{Assumptions.} Zorya assumes correct Ghidra disassembly; execution halts if jumps target unidentified code. We mitigate this via preprocessing and compiler predictable layouts. Currently, Zorya analyzes non-interactive binaries requiring inputs at initialization.

\textbf{Contributions:}
\begin{itemize}
	\item \textbf{Panic-gated concolic execution}: A filtering cascade addressing Go-specific complexity (C1) and scalability (C2) through precomputed reachability, constraint context filtering, and AST-based pre-checking, reducing SMT solver queries by two orders of magnitude. Zorya is open source at \textit{\url{https://github.com/Ledger-Donjon/zorya}}.
	\item \textbf{Empirical validation}: Evaluation on five Go vulnerabilities demonstrates 1.2–3.9× speedups via optimization, with 5/5 detection versus 0–2/5 for existing tools. Evaluation results can be found at \textit{\url{https://github.com/Ledger-Donjon/zorya-evaluation}}.
	\item \textbf{Go vulnerability corpus}: Initial dataset enabling reproducible research at \textit{\url{https://github.com/Ledger-Donjon/logic_bombs_go}}.
\end{itemize}

\section{Background}

\subsection{P-Code Intermediate Representation}
Ghidra's P-Code is a platform-independent intermediate language that normalizes diverse instruction sets into a consistent representation. Each native instruction decomposes into one or more P-Code operations (e.g., \texttt{LOAD}, \texttt{STORE}, \texttt{INT\_ADD}, \texttt{CBranch}). P-Code uses varnodes to represent storage locations: registers, memory addresses, constants, and temporary values \cite{naus_formal_2023}. 

Ghidra distinguishes between raw P-Code, which directly translates machine instructions preserving all low-level details and temporary registers, and high-level P-Code, which applies simplifications for decompilation \cite{eagle_ghidra_2020}. Zorya uses raw P-Code to maintain instruction-level precision required for concolic execution. This normalization facilitates reusing the same concolic infrastructure across x86-64, ARM, and other architectures, while only requiring architecture-specific P-Code lifters and syscall models rather than a full reimplementation of instruction semantics per platform.

\subsection{Symbolic and Concolic Execution}
Symbolic execution analyzes programs by representing inputs as symbolic variables instead of fixed values. When the program reaches a conditional branch, a symbolic executor uses an SMT (Satis\-fiability Modulo Theories) solver to check if alternative execution paths are possible. SMT solvers such as Z3 \cite{de_moura_z3_2008} handle constraints involving integers, bit-vectors, arrays, and other data types, generating concrete input values that satisfy logical conditions. However, symbolic execution faces a fundamental scalability problem: path explosion. Each conditional branch can double the number of paths to analyze, quickly making complete program analysis impractical.

Concolic execution, combining concrete and symbolic execution, mitigates this problem. This approach maintains two parallel states: a concrete state that executes the program with real values, and a symbolic state that records logical constraints on inputs \cite{sen_concolic_2006}. The concrete execution selects a specific path through the program based on the binary's inputs, while the symbolic state captures conditions needed to reach alternative paths. This hybrid strategy improves scalability by following one concrete path at a time while using symbolic reasoning to discover inputs that trigger different behaviors. 

\subsection{Go language and compilers}

\subsubsection{Go components}
Go, introduced in 2009, is a statically typed, compiled language designed for simple concurrent programming \cite{google_go_2025}. Concurrency is a core feature: lightweight goroutines are started with the \texttt{go} keyword and communicate primarily through typed channels, while also supporting shared-memory synchronization when needed \cite{google_effective_2025}. The language provides structured types such as slices and maps; slices are descriptors storing a pointer to the underlying array, a length, and a capacity, and out-of-bounds accesses trigger runtime panics due to mandatory bounds checking \cite{google_go_2025-1}. Go inserts runtime checks to prevent many invalid memory accesses, but it is not fully memory safe: the \texttt{unsafe} package, cgo, and low-level runtime code can still cause segmentation faults and other memory-safety violations \cite{google_go_2025-2}. We therefore focus first on detecting calls to the \texttt{panic} mechanism, which captures most well-structured failures in Go and TinyGo binaries.

\subsubsection{Go compiler}
The standard Go toolchain produces statically linked binaries that bundle the runtime, including the scheduler, garbage collector (GC), and type information, into each executable \cite{google_go_2025}. At startup, execution enters the runtime, which initializes the scheduler, GC, and global state before invoking the user-defined \texttt{main} function, interleaving user code with runtime services throughout the binary. The compiler emits DWARF debug information describing variable locations, types, and source mappings, which we use to reconstruct function arguments and high-level types in our binary-level analysis.

\subsubsection{TinyGo compiler}
TinyGo is an alternative Go toolchain for microcontrollers and other resource-constrained platforms \cite{tinygo-org_documentation_2025}. It introduces a new compiler, which uses (mostly) the standard Go libraries and LLVM to generate machine code, and a lightweight runtime that implements a memory allocator, scheduler, string ope\-rations, and partially re-implemented packages such as \texttt{sync} and \texttt{reflect}. This design avoids limitations of the standard toolchain on microcontrollers, including missing support for instruction sets such as Thumb and AVR and large runtime footprints \cite{tinygo-org_documentation_2025}. By default, TinyGo executes Go code on a single operating system thread and multiplexes goroutines in user space, which removes many low-level thread interleavings and makes TinyGo binaries a simpler first target for symbolic and concolic analysis. Many safety checks, such as slice bounds and nil dereferences, still lead to panics, while TinyGo’s documentation notes that stack overflows and low-level bugs can still cause segmentation faults and hard crashes \cite{tinygo-org_documentation_2025}. Our prototype therefore targets panic-induced failures in TinyGo binaries, leaving non-panic crashes such as segmentation faults to future work.

\section{Motivating Example}
We illustrate Zorya's capabilities on a TinyGo-compiled calculator that accepts two operands and an operator. The aim of using the TinyGo compiler is to work with single-threaded binaries, easier to analyze symbolically, as a first approach. This example program (Listing~\ref{lst:coreEngine}) is correct on typical inputs (e.g., \texttt{2 + 3}), but contains an injected, state-dependent bug: when both operands equal 5, a nil-pointer dereference causes a runtime panic (line 10). Using Zorya's binary-argument symbolic exploration, command-line arguments are modeled symbolically at the starting address, and execution proceeds concolically over P-Code, maintaining both concrete values and symbolic expressions, while building the symbolic path predicate. Upon encountering relevant control-flow splits, Zorya queries an SMT solver to synthesize inputs that satisfy the path conditions leading to the panic site; in this case, it recovers the precise trigger \texttt{operand1 == 5 \&\& operand2 == 5}.

\begin{lstlisting}[language=Go, caption={Core function with injected panic.}, label={lst:coreEngine}]
func coreEngine(num1 int, operator string, num2 int) (int, error) {
    var result int
    switch operator {
    case "+": result = num1 + num2
    case "-": result = num1 - num2
    case "*": result = num1 * num2
    [...]
    // Intentional panic trigger
    if num1 == 5 && num2 == 5 {
        var p *int; *p = 0
    }
    return result, nil
}
\end{lstlisting}

\section{Concolic Execution and Path Predicate Collection}
\label{sec:constraints}
Zorya extracts solver-ready predicates from conditional flags at each P-Code conditional branching (CBranch instruction). Let $\Pi$ denote the \emph{path predicate}, the conjunction of branch conditions along the concrete path, and $\phi$ the symbolic predicate from the branch flag. On traversing an edge, Zorya updates $\Pi'=\Pi \cup \{\phi\}$, accumulating only execution-induced constraints. Concrete values guide the primary path while symbolic expressions enable alternative branch exploration.

Predicates are SMT booleans and bit-vectors, with bit-vectors interpreted as true if and only if non-zero. Negation follows standard semantics. For example, if a guard simplifies to the SMT-LIB expression $(\text{=}\ \texttt{len!142}\ \#x01)$, Zorya asserts the symbolic length equals 1. This instruction-driven accumulation preserves concrete-run fidelity while producing precise, solver-suitable constraints.

\section{Negated-path Exploration}
\begin{figure}[htbp]
\centerline{\includegraphics[scale=0.37]{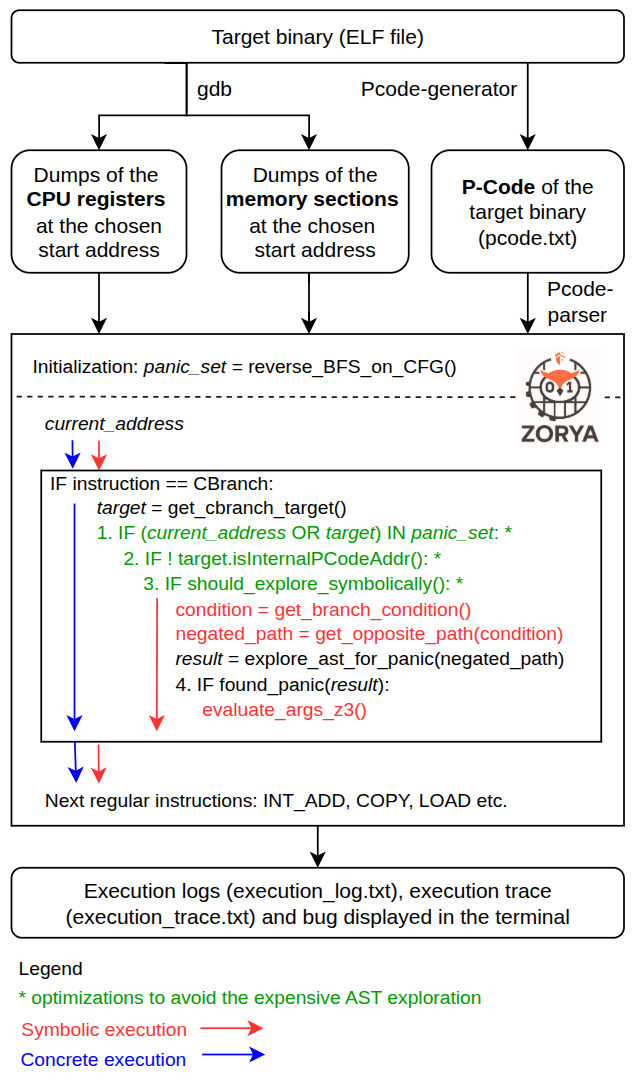}}
\caption{Overview of Zorya workflow, including the optimizations of the negated-path exploration}
\label{fig}
\end{figure}

As illustrated in the Figure \ref{fig}, Zorya performs targeted \emph{negated-path exploration} to synthesize vulnerability-triggering inputs through a panic-gated approach: only branches toward known panic addresses $\mathcal{P}$ undergo symbolic exploration. At each CBranch with target $t$, Zorya applies $\mathrm{IsPanicXref}(t,\mathcal{P})$. For gated branches with predicate $\phi$, the framework snapshots solver state (\texttt{push}), asserts $\Pi'=\Pi\cup\{\neg\phi\}$, queries satisfiability, extracts models on \textsf{SAT}, and discards on \textsf{UNSAT} (\texttt{pop}). Exploration proceeds when $\phi$ references symbolic arguments \emph{or} when prior constraints exist, capturing transitive symbolic dependencies.

To achieve scalability, Zorya implements cascaded filtering that reduces solver invocations through four complementary optimizations applied sequentially at each CBranch. Each stage refines the candidate set, with later stages performing more expensive analysis on progressively smaller subsets:

\textbf{1./ Precomputed Panic Reachability (coarse filter).} Zorya performs static backward analysis from panic-fatal-abort sites to construct $\mathcal{R}$, the set of panic-reachable basic blocks. The algorithm starts from known panic function addresses (identified via symbol analysis of \texttt{runtime.panic*}, \texttt{runtime.fatal*}, etc.) and traverses the control-flow graph backwards, marking all blocks that can reach these sites through any path. This includes both direct calls and indirect control flow through function pointers or virtual dispatch. The analysis is conservative: if uncertainty exists about reachability, blocks are marked reachable, ensuring soundness. At runtime, branches where $pc \notin \mathcal{R}$ and $t \notin (\mathcal{R} \cup \mathcal{P})$ are immediately discarded without symbolic analysis. 

\textbf{2./ Internal Target Detection (artifact filter).} P-Code translation generates internal branches targeting other P-Code statements in the same assembly instruction. Contrary to usual branches, Ghidra represents these targets in the \texttt{const} space, enabling Zorya to detect and exclude them from symbolic analysis. These artifacts arise from complex instruction semantics (e.g., x86 conditional moves) that decompile into conditional P-Code micro-operations never representing user-level control flow.

\textbf{3./ Constraint Context Filtering (relevance filter).} Exploration requires either (i) $\phi$ syntactically references tracked arguments, or (ii) non-empty solver constraints from prior branches, focusing on symbolically-relevant paths. This filter recognizes that branches independent of symbolic inputs cannot contribute to vulnerability discovery: if a branch depends only on constants or non-symbolic state, no input modification can alter its behavior.

\textbf{4./ AST Pre-Check (speculative filter).} Before SMT invocation, lightweight AST traversal from $addr_{alt}$ searches for panic calls within bounded depth (10 blocks). Only \textsf{FOUND\_PANIC} results proceed to solving. This speculative exploration executes symbolically without solver queries, simulating execution along the negated path to determine whether a panic is reachable. The bounded depth prevents unbounded exploration while capturing panic calls within typical inlined function bodies.

\textbf{Single-Evaluation Commitment (efficiency guarantee).} Upon AST confirmation, exactly one solver query executes, minimizing overhead via deferred SMT until high-confidence targets emerge. Traditional symbolic execution may query the solver at every branch; Zorya's deferred approach batches symbolic reasoning until panic proximity is confirmed, concentrating solver invocations on high-value targets.

These optimizations reduce solver queries by up to two orders of magnitude (Section~\ref{sec:eval}). Conceptually, panic-gated exploration instantiates classical target-directed reachability: panic sites play the role of assertions, and the precomputed panic-reachability set over-approximates the basic blocks that can flow to these targets. In contrast to Backward-Bounded DSE (BB-DSE), which uses backward symbolic reasoning to answer \emph{infeasibility} questions on obfuscated x86 code \cite{bardin_backward-bounded_2017}, Zorya combines a lightweight backward control-flow reachability analysis with forward concolic execution to answer \emph{feasibility} questions for panic sites in Go/TinyGo binaries, guided by Go-specific modeling and the multi-layer filter cascade.

\section{Algorithm}
\subsection{Structure}
Algorithm~\ref{alg:zorya} implements Zorya's optimized concolic execution through three phases. The \textbf{initialization phase} (lines 2-11) establishes the analysis infrastructure: lines 2-4 create the Z3 context and solver, while lines 5-7 extract essential metadata from the binary (function signatures via DWARF, panic addresses via symbol analysis, and the precomputed panic-reachability set $\mathcal{R}$). Lines 8-11 initialize symbolic arguments by creating symbolic variables for each function parameter and writing them to their designated locations in the CPU state $\Sigma$.

The \textbf{main execution loop} (lines 12-32) processes P-Code instructions sequentially. For each instruction, the algorithm first checks if it represents a conditional branch (line 16). Non-branch instructions proceed directly to execution (line 34). For conditional branches, the four-layer optimization cascade applies: line 17 implements \emph{internal target detection}, filtering out P-Code internal addresses; line 18 applies \emph{reachability gating}, discarding branches where neither the current location $pc$ nor target $t$ are panic-reachable; line 21 enforces \emph{constraint context filtering}, requiring that the branch condition $\phi$ references symbolic arguments or that prior constraints exist; and line 22 executes \emph{AST pre-checking}, speculatively exploring the negated path to confirm panic reachability before expensive SMT invocation.

The \textbf{SMT solving phase} (lines 23-27) executes only when all filters pass. The solver snapshots its state (line 23), asserts the negated condition $\neg\phi$ (line 24), and checks satisfiability (line 25). On SAT, the algorithm returns a model—concrete input values that would follow the negated path to the panic. On UNSAT, the solver restores its previous state (line 27) and continues execution. This structure implements \emph{single-evaluation commitment}: exactly one solver query per confirmed panic-relevant branch.

\begin{algorithm}
\caption{Zorya: Optimized Concolic Execution}
\label{alg:zorya}
\begin{algorithmic}[1]
\small
\Procedure{ZoryaConcolic}{$binary, start\_addr$}
    \State $\mathcal{C} \gets$ Z3Context()
    \State $\mathcal{S} \gets$ Z3Solver($\mathcal{C}$)
    \State $\Sigma \gets$ InitializeCPUState($\mathcal{C}$)
    \State $\mathcal{F} \gets$ ExtractFunctionSigs($binary$)
    \State $\mathcal{P} \gets$ LoadPanicAddresses($binary$)
    \State $\mathcal{R} \gets$ PrecomputePanicReach($\mathcal{P}$)
    \For{$arg \in \mathcal{F}$.target.arguments}
        \State $\alpha \gets$ CreateSymbolicVar($arg$, $\mathcal{C}$)
        \State WriteToLocation($arg$.reg, $\alpha$, $\Sigma$)
    \EndFor
    \State $pc \gets start\_addr$
    \While{$pc \neq \bot$}
        \State $inst \gets$ FetchInstruction($pc$)
        \State $t \gets \mathrm{TargetOf}(inst)$
        \If{$inst=$CBranch}
            \If{$\neg$IsInternalPCodeAddr($t$)} \Comment{Skip sub-instr}
                \If{$pc \in \mathcal{R} \lor t \in (\mathcal{R} \cup \mathcal{P})$}
                    \State $\phi \gets$ ExtractCondition($inst$)
                    \State $addr_{alt} \gets$ NegatedPath($inst$, $pc$)
                    \If{$\phi$ refs $\mathcal{F}$.args $\lor$ $\mathcal{S}$ has constraints}
                        \If{ExploreAST($addr_{alt}$)=PANIC}
                            \State $\mathcal{S}$.push()
                            \State $\mathcal{S}$.assert($\neg \phi$)
                            \If{$\mathcal{S}$.check() $= $ SAT}
                                \State \Return GetModel($\mathcal{S}$)
                            \EndIf
                            \State $\mathcal{S}$.pop()
                        \EndIf
                    \EndIf
                \EndIf
            \EndIf
        \EndIf
        \State $(\Sigma, pc) \gets$ Execute($inst$, $\Sigma$, $pc$)
    \EndWhile
\EndProcedure
\end{algorithmic}
\end{algorithm}

\subsection{Correctness, Soundness and Completeness}
Zorya is \textit{correct along explored paths}: for each concrete execution it follows, the symbolic state is kept consistent with the concrete state, so the path predicate $\Pi$ faithfully encodes encountered branch conditions. It is \textit{sound with respect to path feasibility} under ideal assumptions on Z3, P-Code semantics, DWARF information, the memory model, and absence of unmodeled nondeterminism: any model $\mathcal{M}$ satisfying $\Pi \cup \{\neg \phi\}$ should then drive execution along the corresponding negated path to the targeted panic. In practice, engineering choices such as \emph{lazy concretization} of symbolic values used in memory accesses, bounded materialization of slices/strings, and pointer anchoring introduce approximations that can miss feasible paths or admit spurious ones in corner cases. As a result, Zorya is \emph{boundedly complete}: these heuristics and the panic-gated search strategy improve scalability at the expense of full soundness over all possible executions.

\section{Implementation}

Zorya's implementation addresses three key challenges: bridging P-Code to executable semantics, reconstructing Go's data layout at binary level, and managing SMT solver efficiency.

\textbf{P-Code Interpretation.} The Rust engine maps varnodes to paired concrete/symbolic values, with on-demand conversion for registers, memory, and constants. A critical challenge is P-Code's temporary registers, which persist across instructions but lack symbolic identity, Zorya tracks these via instruction handling.

\textbf{Go Data Reconstruction.} Without source types, Zorya recovers structures through DWARF and runtime inspection. Slices materialize as $(\texttt{ptr}, \texttt{len}, \texttt{cap})$ triples where pointers are \emph{anchored} to concrete addresses (\texttt{ptr == 0xNNNN}) to prevent symbolic pointer arithmetic bottlenecks, while length/capacity remain symbolic within bounds $[0, 64]$. Strings follow similar patterns (256-byte limit). Zorya relies on DWARF debug information to extract function signatures and argument locations; binaries compiled without debug symbols are not currently supported.

\textbf{Z3 Optimize Solver.} Zorya uses Z3's \texttt{Optimize} solver to handle symbolic constraints. For symbolic slices, Zorya enforces hard constraints on length and capacity (\texttt{0} $\leq$ \texttt{len} $\leq$ \texttt{64}) to maintain tractability while permitting the solver to explore feasible values within these bounds. Push$/$pop scoping maintains solver state across negated-path exploration queries.

\textbf{Concolic Synchronization.} Operations like array/map indexing with symbolic values (\texttt{slice[symbolic\_index]}) require concrete evaluation of \texttt{symbolic\_index} for memory access while retaining symbolic expressions for constraint solving. Zorya employs "lazy concretization": when a symbolic value must be used concretely (e.g., as a memory offset), the framework simplifies the symbolic expression and evaluates it under the current model to obtain a concrete value, while preserving the original symbolic expression for later SMT queries. This approach is critical for Go's runtime, which performs extensive bounds checking with nested conditional expressions that would lead to exponential symbolic expression growth under pure symbolic propagation.

\section{Evaluation}
\label{sec:eval}
We evaluate Zorya's panic-gated optimizations on theoretical and real-world Go binaries, measuring detection accuracy, optimization impact, and performance scaling, under the following research questions:

\begin{itemize}
    \item \textbf{RQ1}: How effectively does Zorya detect vulnerabilities in Go binaries compared to existing symbolic execution tools?
    \item \textbf{RQ2}: What performance improvements do panic-gating and multi-layer filtering provide across different vulnerability types?
    \item \textbf{RQ3}: How does starting point selection (main vs. function) interact with optimization effectiveness?
\end{itemize}

\subsection{Experimental Setup}

Experiments ran on 64-bit Linux with Intel Core i7-1165G7 (4 cores/8 threads, 2.8 GHz base, 4.7 GHz boost) and 32 GB RAM, using Zorya v0.0.4, Ghidra v11.1.4, and TinyGo v0.33.0. We measured average detection time over 5 runs per seed input, comparing unoptimized Zorya (all CBranch explored) against optimized Zorya (multi-layer filtering enabled). We evaluated against BINSEC v0.10.1 \cite{cea_binsecbinsec_2025}, MIASM v0.1.3 \cite{cea-sec_cea-secmiasm_2025}, radius2 v1.0.16 \cite{radare2-org_radare2_2024}, and Owi (verison from on Aug 29, 2025) \cite{ocamlpro_ocamlproowi_2025} with default configurations. 

\subsection{Benchmark}

Our suite combines theoretical programs isolating specific runtime failures with a real-world vulnerability from production audits. The structure of this benchmark is inspired by Logic Bomb \cite{xu_concolic_2017} work with focused Go programs:

\paragraph{Theoretical programs.}
\begin{itemize}
    \item \texttt{crashme} (nil map assignment): conditional dereference after byte check, stressing pointer reasoning.
    \item \texttt{invalid-shift} (buffer overflow): byte-as-index into fixed array, exercising ASCII constraint reasoning.
    \item \texttt{panic-index} (index out-of-bounds): numeric argument guarding slice access, isolating bounds-check logic.
    \item \texttt{broken-calculator} (nil dereference): arithmetic operation with nested control flow triggering dereference on specific operand pairs.
\end{itemize}

\paragraph{Real-world case: Omni Network.} \texttt{omni-vuln} reproduces "OMP-12: Index Out Of Bounds Panic in \texttt{GetMultiProof()}" from Omni Network's audit \cite{sigma_prime_omni_2024}. Malformed Merkle trees with single-child internal nodes cause sibling index \(s\) to exceed \texttt{len(tree)-1}, triggering runtime panic on \texttt{tree[s]} access.

All binaries are single-threaded TinyGo compilations. We pa\-ckage these binaries and their triggering/non-triggering inputs as a Go vulnerability corpus called \textit{logic\_bombs\_go}, released as an accompanying artefact at \textit{\url{https://github.com/Ledger-Donjon/logic_bombs_go}}, to support reproducible evaluation and future work on Go binary analysis.

\subsection{Results and Analysis}

Table~\ref{tab:symbolic_execution_tools} shows that panic-gating optimizations provide complexity-dependent speedups. The "NA" (Not Adapted) designation indicates tools currently lack the CPU instruction and syscall support required to analyze Go binaries, reflecting engineering priorities rather than fundamental limitations.

\textbf{RQ1 - Detection Effectiveness.}

Zorya detected all five vulnerabilities. Among comparison tools, BINSEC, a well-established binary-level symbolic executor with proven effectiveness on C/C++ programs, possesses sufficient x86-64 instruction coverage to attempt Go binary analysis. It detected \texttt{crashme} and \texttt{invalid-shift} in 1 second each, demonstrating efficient performance on programs with shallow control flow. BINSEC did not complete analysis for the three more complex cases (\texttt{panic-index}, \texttt{broken-calculator}, \texttt{omni-vuln}), which involve Go's runtime structures and Go-specific syscalls.

MIASM, radius2, and Owi are marked "NA" as they currently lack prerequisite support for Go binaries. These are capable tools in their target domains; MIASM and radius2's intermediate languages (IL and ESIL) were designed primarily for other architectures and do not yet model the specific x86-64 instruction sequences and system call patterns emitted by Go compilers. Owi was designed for WebAssembly symbolic interpretation and operates on a different instruction set architecture than native x86-64 Go binaries. Extending these tools to support Go would require architectural additions—instruction semantic rules, syscall handlers, and calling convention models, representing reasonable engineering investments that maintainers may pursue based on community needs. 

\smallskip

\noindent
\colorbox{gray!10}{%
    \parbox{0.47\textwidth}{
        \textbf{Finding 1:} Zorya detected 5/5 vulnerabilities through Go-specific modeling. BINSEC detected 2/5 simpler cases; MIASM/radius2/Owi do not yet support Go binary analysis (lacking CPU instruction and syscall models), highlighting the value of language-aware binary analysis.
    }
}

\smallskip

\textbf{RQ2 - Optimization Impact.}

Multi-layer filtering provides speedups that vary significantly based on whether panic-reachability gating filters branches. When gating statistics show [0/N], meaning zero branches filtered—results are mixed: \texttt{crashme} improves from 45.1s to 35.6s (1.3× speedup) in main mode but degrades from 23.6s to 35.8s (0.7× slowdown) in function mode, while \texttt{invalid-shift} improves 1.5× (main) and 1.4× (function). These programs have shallow control flow where every branch is panic-reachable, so only the four non-gating optimizations apply. The inconsistent results suggest that optimization overhead (tracking gating statistics, AST pre-checks) can outweigh benefits when no branches are actually filtered.

When panic-reachability gating actively filters branches, results are consistently positive. \texttt{panic-index} filters 33–47\% of branches and achieves 2.4–3.9× speedups; \texttt{broken-calculator} filters 35–70\% and gains 1.8–2.2×; \texttt{omni-vuln} filters 38\% and improves 3.3×. Here, the gating optimization eliminates entire branches before other stages process them. For example, \texttt{panic-index} in function mode gates 8 of 17 branches (47\%) and achieves 3.9× speedup, while main mode gates only 19 of 58 branches (33\%) and achieves 2.4× speedup—suggesting that gating effectiveness matters more than the absolute number of branches filtered.

Notably, \texttt{omni-vuln} transforms from intractable ($>$7200s) to practical (83.1s) despite filtering only 38\% of branches, demonstrating that even moderate filtering can enable analysis of complex real-world programs. The data reveals a threshold effect: programs where gating filters zero branches see marginal or negative results due to the additional calculation of the reverse BFS (0.7–1.5×), while programs where gating filters any branches see substantial gains (1.8–3.9×). This confirms that panic-reachability gating is the critical optimization: when it activates, the four-layer cascade delivers strong performance. When it cannot filter branches, the remaining optimizations provide limited benefit or even introduce overhead.

\smallskip
\noindent
\colorbox{gray!10}{%
    \parbox{0.47\textwidth}{
        \textbf{Finding 2:} Optimizations provide 1.8–3.9× speedups when panic-reachability gating filters branches (33–70\% filtering). When no branches are filtered, results are inconsistent (0.7–1.5×), with one case showing slowdown, indicating that gating effectiveness is critical for optimization benefit.
    }
}
\smallskip

\textbf{RQ3 - Starting Point and Optimization Interaction.}

Function mode benefits more from optimizations than main mode. For \texttt{panic-index}, function mode achieves 3.9× speedup compared to main mode's 2.4×. This reflects differing baseline costs: main mode includes Go runtime initialization overhead orthogonal to panic reachability, which moderates optimization impact. Function mode bypasses this infrastructure, concentrating on user logic where panic-gating more directly reduces exploration overhead.

Simpler programs (\texttt{crashme}, \texttt{invalid-shift}) show convergence after optimization (35–41s), suggesting both modes reach similar bottlenecks once filtering addresses shared overhead. The real-world \texttt{omni-vuln} illustrates a practical consideration: main mode remains intractable ($>$7200s) due to CLI parsing complexity, while function mode becomes viable (83.1s). 

\begin{table*}[t]
\caption{Evaluation of Zorya and other tools on theoretical and real-world vulnerabilities.}
\label{tab:symbolic_execution_tools}
\centering
\small
\setlength{\tabcolsep}{5pt}
\renewcommand{\arraystretch}{1.2}
\begin{tabular}{@{}lclllrrrcc@{}}
\toprule
\textbf{Binary} & \textbf{Size} & \textbf{Tested inputs} & \textbf{Vuln.} & \textbf{Starting} & \textbf{Zorya before} & \textbf{Zorya after} & \textbf{Gated/} & \textbf{BINSEC} & \textbf{Other} \\
\textbf{(Type)} & \textbf{(KB)} & \textbf{(no panic)} & \textbf{input(s)} & \textbf{point} & \textbf{opt. (sec)} & \textbf{opt. (sec)} & \textbf{Total} & \textbf{(sec)} & \textbf{tools} \\
\midrule
crashme & 866 & "a"; "B"; "100" & "K" & main.main & 45.1 & \textbf{35.6} & 0/4 & 1 & NA \\
(Nil Map) &  &  &  & crash() & 23.6 & \textbf{35.8} & 0/1 &  &  \\
\midrule
invalid-shift & 866 & "10"; "42"; "1000" & "@"; "H?"; & main.main & 53.5 & \textbf{36.2} & 0/5 & 1 & NA \\
(Buffer overflow) &  &  & "d???" & shift() & 55.9 & \textbf{40.6} & 0/2 &  &  \\
\midrule
panic-index & 885 & "0"; "1"; "2" & arg1 $>$ 3 & main.main & 489.9 & \textbf{203.1} & 19/58 & NA & NA \\
(Index OOB) &  &  &  & index() & 148.3 & \textbf{38.3} & 8/17 &  &  \\
\midrule
broken-calc & 930 & "2+3"; "5+1"; & arg1=5 & main.main & 726.6 & \textbf{331.1} & 34/97 & NA & NA \\
(Nil Dereference) &  & "6-5" & arg3=5 & coreEngine() & 77.6 & \textbf{44.1} & 7/10 &  &  \\
\midrule
omni-vuln & 1403 & "a b c [e f [g h]] & "0 0 & main.main & $>$7200 & $>$7200 & -- & NA & NA \\
(Merkle tree) &  & --indices [1$|$3$|$5]" & --indices 1" & \mbox{GetMultiProof()} & 273.1 & \textbf{83.1} & 12/32 &  &  \\
\bottomrule
\end{tabular}
\vspace{0.5em}
\par\noindent\scriptsize \textbf{Abbreviations:} OOB = Out-of-Bounds; Before/After opt. = Before/After adding the optimizations to Zorya; Gated/Total = number of CBranch instructions gated by panic reachability out of total; Other tools = MIASM, radius2, Owi; NA: Not Adapted for Go binaries due to lack of support; Note: Average detection times are calculated with the inputs that do not trigger panics, over five runs each. 
\end{table*}

This difference of at least 87× indicates that function-level analysis may be preferable for certain deployment scenarios. This divergence reveals a natural trade-off in concolic execution design: main-mode analysis offers broader coverage by capturing argument parsing logic and environment interactions, but experiences path explosion in framework code (CLI parsers, logging initializers) unrelated to core vulnerability logic. Function-mode analysis trades some of this coverage for improved tractability, assuming developers can identify candidate vulnerable functions, a reasonable assumption for targeted security audits, though potentially less suitable for whole-program bug discovery. This suggests that a hybrid approach, where lightweight static analysis first identifies high-risk functions for deeper function-mode exploration, could combine the benefits of both strategies.

\smallskip

\noindent
\colorbox{gray!10}{%
    \parbox{0.47\textwidth}{
        \textbf{Finding 3:} Function-mode analysis amplifies optimization benefits, achieving 1.8–3.9× speedups by concentrating filtering on user logic. Main mode's initialization overhead makes it less tractable for complex real-world cases.
    }
}

\subsection{Discussion}

The correlation between gating statistics and speedup (0\% gating/1.3× to 70\% gating/3.9×) provides empirical support for multi-layer filtering. Programs with deep control flow benefit most from panic-reach precomputation, which prunes 35–70\% of branches before symbolic analysis. Programs with shallow control flow see more modest gains, as expected when all branches are panic-relevant.

BINSEC's success on simpler cases demonstrates that general-purpose symbolic executors with comprehensive x86-64 support can effectively analyze Go binaries when execution remains relatively shallow. The challenges observed with more complex cases suggest that Go's compiled runtime introduces semantic layers that may benefit from language-specific modeling. Tools marked "NA" do not yet include the foundational instruction and syscall support needed for Go analysis, a gap that could be addressed through targeted extensions if tool maintainers identify sufficient demand from their user communities. Zorya addresses these requirements through P-Code normalization and Go-aware symbolic types, offering one approach to binary-level Go vulnerability discovery that may complement existing tools' strengths in other domains.

\section{Limits and Improvements}
Zorya has several areas for future enhancement. Currently, Zorya operates on non-interactive binaries, requiring all inputs at program initialization (command-line arguments or function parameters). This design choice simplifies the initial implementation but limits analysis of programs with runtime user interaction, file I/O loops, or network-driven input processing. Extending Zorya to handle incremental symbolic input during execution would broaden its applicability to interactive applications such as servers, REPLs, and streaming processors.

The panic-reachability analysis could be enhanced to eliminate the AST pre-check phase. Currently, the reverse BFS produces a simple set enabling constant-time membership queries. Augmenting this set with path information during preprocessing could avoid runtime AST exploration, but risks complicating the data structure and degrading lookup performance. We maintain the current two-phase design (reverse BFS + AST pre-check) to preserve linear-time panic-reachability queries, deferring finer-grained analysis to runtime when high-confidence targets emerge.

Additional improvements could include integrating fuzzing techniques to reduce reliance on concrete seed inputs for initial path exploration, dynamically adjusting AST exploration depth based on program complexity, and reconsidering the exclusion of internal Go functions from negated-path exploration to potentially capture runtime vulnerabilities. 

Extending Zorya to multi-threaded Go binaries represents a significant research challenge. Go's concurrency model, based on goroutines and channels, introduces interleaving explosion: each synchronization point multiplies the exploration space exponentially. Concurrency-specific panics, such as sending on closed channels, unsynchronized map access, or deadlocks, require symbolically modeling Go's runtime scheduler and synchronization primitives. Despite these challenges, supporting concurrent programs would significantly expand Zorya's applicability, as goroutines pervade real-world Go applications where concurrency bugs often manifest as panics under race conditions.

\section{Related Work}

Binary-level symbolic execution tools have trouble with Go binaries because of the Go runtime and data layout. LLVM IR-based tools such as Haybale \cite{plsyssec_haybale_2024} and SymSan \cite{r-fuzz_symsan_2024} depend on gollvm \cite{go-community_gollvm_2017}, which is not widely used in production and does not yet support many real Go builds. Even when IR is available, Go features (slices as $\langle$ptr,len,cap$\rangle$, interfaces as $\langle$type,value$\rangle$, channels) are lowered to generic memory operations, which hides the checks and invariants needed for precise constraints. MAAT \cite{trail_of_bits_maat_2024} also uses Ghidra's P-Code like Zorya, but lacks Go-specific semantic modeling and cannot direct solving toward panic-relevant paths. In our evaluation, MAAT reported unsupported instruction errors on Go binaries, preventing successful analysis.

General-purpose binary analyzers such as Angr \cite{wang_angr_2017} and MIASM need Go-specific lifters, syscall models, and runtime stubs; without them they struggle with garbage collection and calling conventions. Radius2, built on radare2’s ESIL, uses a lightweight stack-based IR for emulation and analysis. ESIL does GhiHorn \cite{cert_coordination_center_ghihorn_2021} translates P-Code to Horn clauses for proofs, which is hard to scale to large Go programs and is not focused on input generation. Owi targets WebAssembly and does not analyze native x86-64 Go binaries. BINSEC is first and foremost a general binary analysis framework (disassembly, taint, abstract interpretation), which also includes a symbolic execution engine. It provides strong x86-64 coverage and efficient path exploration on binaries targeting C-like ABIs. However, it does not model Go’s runtime (slices, interfaces, channels, panic routines) or Go-specific syscalls. 

Targeted symbolic execution and line/target reachability have been studied extensively. Backward-Bounded DSE (BB-DSE) \cite{bardin_backward-bounded_2017} performs backward symbolic reasoning from a chosen target to answer \emph{infeasibility} questions on obfuscated x86 code (e.g., proving that a branch is dead). Zorya's panic-gated exploration is complementary: it also focuses on specific targets, but uses conservative backward control-flow reachability followed by forward concolic execution to answer \emph{feasibility} questions for panic sites in Go/TinyGo binaries.

Go-focused work is limited. ColorGo \cite{li_colorgo_2025} proposes directed concolic execution at the source level, but there is no public tool to compare against and it does not address binary-only settings. DuckEEGO \cite{shao_duckee_2018} supports only basic types (e.g., ints, bools, simple maps) and does not handle strings, slices, structs, goroutines, or external libraries, which are common in real programs. Zorya addresses these gaps through a unique combination of P-Code-based analysis, DWARF-based argument recovery, binary-level Go type modeling, and panic-gated exploration, enabling practical vulnerability detection in compiled Go binaries without source code.

\section{Conclusion}

This paper presented several improvements to Zorya, a panic-gated concolic execution framework for Go binaries that combines P-Code lifting with Go-aware symbolic modeling. Evaluation on five vulnera\-bilities demonstrates that multi-layer filtering provides 1.8–3.9× speedups when panic-reachability gating filters 33–70\% of branches, but introduces overhead when no filtering occurs, validating panic-gating as the critical optimization. Zorya detected all five vulnerabilities while comparison tools detected at most two, highlighting the value of language-specific binary analysis.

Function-mode analysis proved essential for real-world deployment, running roughly two orders of magnitude faster than main-mode on complex programs by bypassing runtime initialization overhead. Future work will extend Zorya to multi-threaded binaries and support incremental symbolic inputs. This work establishes that specialized concolic execution can achieve practical vulnerabi\-lity detection in language ecosystems with runtime safety checks.

\section*{Acknowledgment}
The authors would like to thank our anonymous reviewers for their valuable feedback.
This research would not have been possible without the support of the Ledger Donjon team and the Telecom Paris INFRES department. We extend special thanks to Dr. Robin David for his invaluable guidance and advice throughout this work.

\bibliographystyle{ACM-Reference-Format}
\bibliography{sample-base}

\end{document}